\documentclass[a4paper,11pt]{article}

\usepackage{jheppub} 

\usepackage{graphicx}

\usepackage{amsmath}
\usepackage[utf8]{inputenc}

\usepackage{amssymb}
\usepackage{amsfonts}
\usepackage{amsbsy}
\usepackage{url}
\usepackage{enumerate}
\usepackage{caption}
\usepackage{subcaption}
\usepackage{color}
\usepackage{ulem}
\usepackage{bm}
\usepackage{comment}
\usepackage{multirow,bigdelim}

\newcommand{\be}{\begin{eqnarray}}
\newcommand{\ee}{\end{eqnarray}}
\newcommand{\p}{\partial}

\makeatletter
\@addtoreset{equation}{section}
\makeatother

\newcommand{\bee}{\begin{equation}}
\newcommand{\eee}{(\end{equation})}

\newcommand\rsout{\bgroup\markoverwith{\textcolor{red}{\rule[0.5ex]{2pt}{0.4pt}}}\ULon}

\begin{document}

\preprint{YGHP-22-3}

\title{Quantum nucleation of topological solitons}

\author{Minoru Eto$^{1,2}$, Muneto Nitta$^{2,3}$}

\affiliation{%
$^1$Department of Physics, Yamagata University, 
Kojirakawa-machi 1-4-12, Yamagata,
Yamagata 990-8560, Japan, \\
$^2$Research and Education Center for Natural
Sciences, Keio University, Hiyoshi 4-1-1, Yokohama, Kanagawa 223-8521, Japan
 \\
$^3$Department of Physics, Keio University, Hiyoshi 4-1-1, Yokohama, Kanagawa 223-8521, Japan
}%

\date{\today}

\abstract{
The chiral soliton lattice is 
an array of topological solitons 
realized as
ground states 
of QCD at finite density under strong magnetic fields or rapid rotation, 
and chiral magnets with an easy-plane anisotropy.
In such cases, topological solitons have negative energy due to 
topological terms originating from the chiral magnetic or vortical effect  
and 
the Dzyaloshinskii-Moriya interaction, respectively.
We study quantum nucleation of topological solitons 
in the vacuum through quantum tunneling  
in $2+1$ and $3+1$ dimensions,
by using a complex $\phi^4$ (or the axion) model with a topological term proportional to an external field, 
which is a simplification of low-energy theories of the above systems. 
In $2+1$ dimensions, a pair of a vortex and an anti-vortex is connected by a linear soliton,  
while in $3+1$ dimensions, a vortex is string-like, a soliton is wall-like,
and a disk of a soliton wall is bounded by a string loop. 
Since the tension of solitons can be effectively negative due to the topological term, 
such a composite configuration of a finite size is created 
by quantum tunneling and subsequently grows rapidly.
We estimate the nucleation probability analytically 
in the thin-defect approximation and fully calculate it numerically using the relaxation (gradient flow) method.  
The nucleation probability is maximized when the direction of the soliton is perpendicular to the external field. 
We find a good agreement between the thin-defect approximation 
and direct numerical simulation in $2+1$ dimensions  
if we read the vortex tension from the numerics,
while we find a difference between them at short distances 
interpreted as a remnant energy  
 in $3+1$ dimensions.
}

\maketitle


\section{Introduction}
\label{sec:intro}

Topological solitons 
are ubiquitous in nature, 
playing significant roles in 
quantum field theories 
\cite{Rajaraman:1987,Rubakov:2002fi,Manton:2004tk, Shnir:2005vvi,Vachaspati:2006zz,Dunajski:2010zz,Weinberg:2012pjx,Shnir:2018yzp},
supersymmetric gauge theories 
\cite{Tong:2005un,Tong:2008qd,Eto:2006pg,Shifman:2007ce,Shifman:2009zz}, 
QCD 
\cite{Eto:2013hoa},
cosmology 
\cite{Kibble:1976sj,Kibble:1980mv,Vilenkin:1984ib,Hindmarsh:1994re,Vachaspati:2015cma,Vilenkin:2000jqa}, 
and various condensed matter systems 
\cite{Mermin:1979zz}
such as  
helium-3 superfluids \cite{Volovik:2003fe}, 
superfluids \cite{Svistunov:2015}, 
superconductors 
\cite{Blatter:1994zz,Giamarchi:2002}, 
Josephson junctions of superconductors \cite{Ustinov2015}, 
Bose-Einstein condensates (BECs) of ultracold atomic gasses 
\cite{Kawaguchi:2012ii}, 
nonlinear media 
\cite{Pismen,Bunkov:2000}, 
liquid crystals \cite{Pismen}  
and magnets \cite{Nagaosa2013,Gobel:2020mqd}.
Among various salient features of topological solitons, 
formations of topological solitons are 
one of the most important aspects.
Topological solitons are known to be created 
at phase transitions accompanied with 
a spontaneous symmetry breaking (SSB). 
This is known as the Kibble-Zurek mechanism 
\cite{Kibble:1976sj,Kibble:1980mv,Zurek:1985qw,Zurek:1996sj} 
and was experimentally confirmed in various condensed matter systems  
such as 
liquid crystals \cite{Bowick:1992rz},
helium-4 superfluids \cite{Hendry:1994},  
helium-3 superfluids \cite{Baeuerle:1996zz,Ruutu:1995qz}, 
and ultracold atomic gasses 
\cite{Sadler:2006,Weiler:2008}.
In two spatial dimensions, a pair of a vortex and an anti-vortex 
is created at finite temperature, which is known as
Berezinskii-Kosterlitz-Thouless transition (BKT) transitions 
\cite{Berezinsky:1970fr,Berezinsky:1970-2,Kosterlitz:1972,Kosterlitz:1973xp} 
(see also Refs.~\cite{Kobayashi:2018ezm,Kobayashi:2019sus} 
for recent studies).
This was  experimentally confirmed in thin-film 
helium-4 superfluids \cite{PhysRevLett.40.1727}, 
superconductors \cite{PhysRevLett.42.1165},
and ultracold atomic gases \cite{Hadzibabic:2006}.
There are several other mechanisms for creating topological solitons 
such as 
bubble collisions \cite{Hawking:1982ga,Copeland:1996jz,Digal:1995vd}, 
monopole pair creation under strong magnetic field 
(as a dual to the Schwinger mechanism)
\cite{Affleck:1981ag, Affleck:1981bma}, 
and brane annihilations 
\cite{Sen:2004nf} 
(see 
Refs.~\cite{Bradley:2008zza,Takeuchi:2012ee,Nitta:2012hy,Takeuchi:2012ap} 
for condensed matter analogues).

The purpose of this paper is to propose a yet another mechanism for 
a creation of topological solitons, that is,  
quantum nucleation through quantum tunneling.
This mechanism works when 
the ground state is ``solitonic''. 
When the Lagrangian or Hamiltonian 
contains a certain type of 
a topological term 
with its coefficient  
 larger than a certain critical value,  
the energy of topological solitons is negative 
and thus they are spontaneously created in uniform states.
However, one cannot place infinite number of solitons since 
they repel each other, and thus 
the ground state is a lattice of topological solitons. 
A typical example of solitonic ground states is given by 
chiral soliton lattices (CSLs)  
which are periodic arrays of domain walls or solitons, 
appearing 
in various condensed matter systems: 
cholesteric liquid crystals \cite{DEGENNES1968163} 
and 
chiral magnets \cite{
togawa2012chiral,togawa2016symmetry,KISHINE20151,
PhysRevB.97.184303,PhysRevB.65.064433,Ross:2020orc} 
with 
the Dzyaloshinskii-Moriya (DM) interaction \cite{Dzyaloshinskii,Moriya:1960zz}.
The latter has an important nanotechnological application 
in information processing such as magnetic memory storage devices and magnetic sensors \cite{togawa2016symmetry}. 
The $O(3)$ sigma model together with the DM term 
reduces to 
the sine-Gordon model plus a topological term at low energy, 
and the CSL is a sine-Gordon lattice.
Another condensed matter example of solitonic ground states is 
given by 
magnetic skyrmions 
\cite{Bogdanov:1989,Bogdanov:1995} 
in chiral magnets, 
which typically constitute 
 a triangular lattice in the ground state 
 in the parameter region in which the DM term is strong enough 
\cite{Rossler:2006,Han:2010by,Lin:2014ada,Ross:2020hsw}. 
Since they have been realized in laboratory experiments 
\cite{doi:10.1126/science.1166767,doi:10.1038/nature09124},  
there has been great interests such as an application to 
 information carriers in ultradense memory and logic devices  
  with low energy consumption
\cite{doi:10.1038/nnano.2013.29}. 
The other examples are for instance 
${\mathbb C}P^{N-1}$ skyrmion lattices in
$SU(N)$ magnets 
\cite{Akagi:2021dpk, Akagi:2021lva, Zhang:2022, 
Amari:2022boe} and 3D skyrmions 
\cite{Kawakami:2012zw} 
in spin-orbit coupled BECs 
with background gauge fields as generalizations of the DM term.

Recently, it has been predicted  
that CSLs are also ground states 
of QCD at finite density under
strong magnetic field 
\cite{Son:2007ny,Eto:2012qd,Brauner:2016pko,Chen:2021vou,Gronli:2022cri}  
or under rapid rotation 
\cite{Huang:2017pqe,Nishimura:2020odq,Eto:2021gyy}, 
due to a topological term originated from 
the chiral magnetic effect (CME) 
\cite{Son:2004tq,Son:2007ny} 
which is the vector current in the direction of the magnetic field, 
or chiral vortical effect 
(CVE) \cite{Vilenkin:1979ui,Vilenkin:1980zv,
Son:2009tf} which is the axial vector current in the direction of the rotation axis, respectively. 
They also appear with thermal fluctuation \cite{Brauner:2017uiu,Brauner:2017mui,Brauner:2021sci}
(see also Refs.~\cite{Yamada:2021jhy,Brauner:2019aid,Brauner:2019rjg}).
In the CSLs,
the number density of solitons is determined by 
the strength of external fields such as 
a magnetic field or rotation 
(or the DM term for chiral magnets).
As external fields are larger 
 above the critical value,
the soliton number density is larger.
Thus, when one gradually increases(decreases) 
the strength of the external field,
the mean inter-soliton distance decreases(increases) accordingly. 
One of natural questions is 
how they are created from the vacuum (uniform state).
When one instantaneously changes 
the external field from the value below the critical value 
to the one above the critical value,
it is unnatural that a flat soliton (domain wall) 
with infinite world-volume instantly appears. 
Instead, quantum nucleation can occur in this case 
as we propose in this paper.

To explain our mechanism, it is worth to recall  
quantum decay of a metastable false vacuum and bubble nucleation 
first formulated by Coleman
\cite{Coleman:1977py,Callan:1977pt,Coleman:1987rm}
(see Refs.~\cite{Rubakov:2002fi,Weinberg:2012pjx,Coleman:1985rnk}
 as a review).
Decay probabilities can be calculated by 
evaluating the Euclidean action values for
bounce solutions.
In the thin-wall approximation, one can evaluate 
the decay probability in terms of tensions of domain walls.
Preskill and Vilenkin studied 
quantum decays of metastable topological defects
\cite{Preskill:1992ck} 
(see Ref.~\cite{Vilenkin:2000jqa} as a review, 
and Refs.~\cite{Monin:2008mp,Monin:2008xx,Monin:2009gi} 
for recent studies). 
One of typical cases is given by an axion model, 
in which a domain wall (or soliton) terminates on a string. 
Thus, a domain wall is metastable and 
can decay by quantum tunneling with creating a hole 
bounded by a closed string. 
Again in the thin-wall approximation, one can evaluate 
the decay probability of the domain wall 
in terms of tensions of domain walls and strings. 
Some examples are given by domain walls in two-Higgs doublet models 
\cite{Eto:2018hhg,Eto:2018tnk} 
and axial domain wall-vortex composites 
in QCD \cite{Eto:2013bxa}. 
Another case is a string (vortex) ending on a monopole. 
In this case, a string is metastable  and 
decays by cutting the string into two pieces 
whose endpoints are attached by a monopole and an anti-monopole 
through quantum tunneling.
Examples  
can be found 
for instance for  
electroweak $Z$-strings in the standard model 
\cite{Nambu:1977ag,Achucarro:1999it,Eto:2012kb} and 
non-Abelian strings in dense QCD \cite{Vinci:2012mc,Eto:2013hoa}.

In this paper, 
we study quantum nucleation of topological solitons 
through quantum tunneling. 
For definiteness,
we discuss chiral solitons in a complex $\phi^4$ model 
(an axion model with the domain wall number one) with a topological term, 
which is a simplification of  low-energy theories of 
chiral magnets (with an easy-plane anisotropy) 
and QCD at finite density 
under strong magnetic field or rapid rotation. 
The origin of the topological term is 
 the DM interaction for chiral magnats, while it is 
CME and CVE for QCD under strong magnetic field or rapid rotation, 
respectively. 
If the external field $B$ is larger than a certain critical value 
$B_c$, the soliton tension is
effectively negative, and therefore it can be created by quantum tunneling.
We estimate the nucleation probability analytically 
in the thin-defect approximation in any dimension,  
and fully calculate it numerically 
in $2+1$ and $3+1$ dimensions by
using the relaxation (gradient flow) method.
In $2+1$ dimensions, a vortex is particle-like,  
a soliton is string-like,  and 
a pair of a vortex and an anti-vortex is connected 
by a linear soliton,  
while 
in $3+1$ dimensions, a vortex is string-like, a soliton is wall-like,
and a disk of a soliton wall is bounded by a string loop.
Once such a composite configuration of a finite size is created 
by quantum tunneling, it grows rapidly.
The nucleation probability is maximized 
when the direction of the soliton is perpendicular to 
the external field. 
We also find that decay (nucleation) is prohibited for 
$B>B_c$ ($B<B_c$).
We find that 
 the nucleation probabilities 
calculated in the thin-defect approximation 
and in the direct numerical simulations show 
a good agreement in $2+1$ dimensions  
once we read the vortex tension from the numerics.
On the other hand,  in $3+1$ dimensions, we find a difference between them at short distances at the subleading order 
which we interpret as a remnant energy.

This paper is organized as follows.
In Sec.~\ref{sec:decay_wall}, we give a brief review 
of quantum decay of a soliton in the complex $\phi^4$ model (the axion model with the domain wall number one)
without a topological term.
In Sec.~\ref{sec:nucleation_wall}, 
we present our model (the complex $\phi^4$ model with a topological term) 
and discuss 
quantum nucleation and decay probabilities of solitons 
in the thin-defect approximation.
In Sec.~\ref{sec:numerical}
we numerically calculate the creation probabilities of solitons
in $2+1$ and $3+1$ dimensions  
and compare those in the thin-defect approximation.
Section \ref{sec:summary} is devoted to a summary and discussion.
In Appendix \ref{sec:app2}, we present 
an asymptotic behavior of the scalar field outside 
a pair of vortex and an anti-vortex connected by a soliton.
In Appendix \ref{sec:app1}, we give a derivation of some formula 
used in the quantum nucleation.


\section{Quantum decay of solitons by nucleation of holes: a review}
\label{sec:decay_wall}

We start with giving a brief review of the quantum decay of 
solitons (domain walls) 
 by quantum nucleations of holes 
in a complex $\phi^4$ model (an axion model with the domain  wall number one).
The minimal model in $3+1$ dimensions is
\be
{\cal L}_{\rm UV} = \left|\p_\mu \phi\right|^2 - \frac{\lambda}{4}\left(|\phi|^2 - v^2\right)^2 + v m^2(\phi + \phi^*),
\label{eq:Lag_axion}
\ee
with $v$ and $m$ are parameters whose mass dimension is 1, and $\lambda$ is dimensionless.
If the third term in Eq.~(\ref{eq:Lag_axion}) is absent, the model is 
the Goldstone model invariant under a global $U(1)$ transformation $\phi \to e^{i\eta}\phi$
spontaneously broken in the homogeneous vacuum $|\phi| = v$.
There is a Nambu-Goldstone (NG) mode and
a Higgs mode whose mass is $m_h = v\sqrt{\lambda}$.
When we turn on the third term, the $U(1)$ symmetry is explicitly broken, 
leaving the unique vacuum 
where the NG mode becomes a pseudo-NG mode with the mass $m$. 
The $U(1)$ symmetry is 
an approximate symmetry when the mass of the pseudo-NG mode is sufficiently small 
\be
m_h \gg m
\quad \Leftrightarrow \quad m^2 \ll \lambda v^2.
\label{eq:approxU(1)_cond}
\ee
There the vacuum expectation value can be approximated as
$
\left<\phi\right>  \simeq v \left( 1+ \frac{m^2}{m_h^2}\right).
$

The model admits two kinds of solitonic objects, 
namely vortices and solitons (or domain walls).
The vortex is a global string with co-dimension two which is a topological defect if the explicit $U(1)$ breaking term is absent.
Thickness of the string and the tension of the string for $m=0$, namely the energy per unit length, are given by
\be
\delta_{\rm st} \sim m_h^{-1},\qquad 
\mu\big|_{m\to 0} \sim \pi v^2\log (m_h L),
\label{eq:global_string}
\ee
where $L$ is a long distance cutoff.
When the $U(1)$ breaking term is not zero, the string is no longer topological and it is
always accompanied by the soliton which is a wall-like object with co-dimension one. 

Probably the soliton can be most clearly seen in the limit of $\lambda \to \infty$
where the wine bottle potential becomes infinitely steep ($m_h \to \infty$), so that the amplitude of $\phi$ freezes out as $|\phi|=v$.
Writing $\phi = v e^{i\theta}$ and plugging it
into Eq.~(\ref{eq:Lag_axion}), we are lead to the sine-Gordon model
\be
{\cal L}_{\rm IR} = v^2 \left[(\p_\mu\theta)^2 + 2 m^2 (\cos \theta -1)\right],
\label{eq:Lag_sG0}
\ee
where we have subtracted the constant $2m^2v^2$ for the minimum of the potential energy to be 0 for convenience.
The ground state is homogeneous as $\theta = 0$ with the redundancy of $2\pi n$  $(n \in \mathbb{Z})$. 
The ground state energy $E^{({\rm vac})}=0$.
In addition, there is a sine-Gordon soliton 
which we take perpendicular to the $z$-axis 
without loss of generality:
\be
\theta = 4 \tan^{-1}e^{m z}.
\ee
This connects $\theta = 0$ at $z\to-\infty$ and $\theta = 2\pi$ at $z\to\infty$. 
The thickness and the tension, namely the energy per unit area, of the soliton are given by
\be
\delta_{\rm dw} = m^{-1},\qquad
\sigma\big|_{\lambda \to \infty} = 16 m v^2.
\label{eq:sG_tension}
\ee
Note that the soliton in the sine-Gordon limit is classically stable but
it could be quantum-mechanically unstable. 
This is because 
it can end on 
a sting which is infinitely thin 
($\delta_{\rm st} \to 0$) in the $\lambda \to \infty$ limit, so that 
holes surrounded by the strings can be created by quantum tunneling effect.

The instability of the soliton for a finite $\lambda$ is
two fold: 1) classical instability and 2) quantum instability.

1) The soliton in the UV theory is a metastable non-topological soliton. This is because a loop surrounding the $S^1$ vacuum manifold
which slightly slants by $vm^2(\phi + \phi^*) \sim 2 m^2v^2 \cos\theta$
can pass slip the potential barrier around $|\phi|=0$ and shrink to the unique vacuum. 
This is the classical instability of the soliton
in the UV theory with finite $\lambda$. Roughly speaking, 
if the approximate $U(1)$ condition in Eq.~(\ref{eq:approxU(1)_cond}) is satisfied, 
the soliton remains classically meta-stable.
As for the tension of the string, 
it is quite different for the massive case $m\neq 0$ 
from the massless case $m=0$ in which the tension is logarithmically 
divergent as in Eq.~(\ref{eq:global_string}).\footnote{
In Ref.~\cite{Preskill:1992ck},
a string is attached by
a sine-Gordon domain soliton as our case,
but it is assumed that the energy of the vortex 
is logarithmically divergent.} 
The key point is that 
the amplitude of the scalar field 
 converges to the VEV exponentially fast 
 as numerically confirmed in Appendix \ref{sec:app2}, 
 in contrast to the massless case for which 
 the amplitude polynomially approaches to the VEV.
Thus, 
the tension is finite
\be
\mu\big|_{m > 0} = \text{const.} \label{eq:mu}
\ee
 in contrast to the massless case in Eq.~(\ref{eq:global_string}).
This fact is crucial for the nucleation of topological soliton. 
Note that the approximate $U(1)$ condition in Eq.~(\ref{eq:approxU(1)_cond}) implies $\delta_{\rm st} \ll \delta_{\rm dw}$.

2) Even when the soliton is classically metastable, it would quantum mechanically be unstable
because of the nucleation of a hole. 
Let us assume a shape of a hole is circular.
If the radius $R$ of the hole is much greater than the soliton thickness $\delta_{\rm dw}$, 
we can use the thin-defect approximation, providing
the decay probability \cite{Preskill:1992ck}
\be
P_\text{decay} = A e^{-S},\quad S = \frac{16 \pi \mu^3}{3\sigma^2},\quad R = \frac{2\mu}{\sigma}
\ee
where $\mu$ is the constant tension of the string and 
the soliton tension $\sigma$ is well approximated by 
Eq.~(\ref{eq:sG_tension}) if $\lambda \gg 1$.
The thin-defect approximation is justified for $R\sim \mu/\sigma \sim \mu/(mv^2) \gg 1/m$,
namely $\mu/v^2 \gg 1$.
Thus, the value of the bounce action $S$ 
is justified in the thin defect approximation if 
$S \sim
\mu^3/\sigma^2 \sim \mu^3/(m^2v^4) = (v^2/m^2)(\mu/v^2)^3 \gg v^2/m^2$.
On the other hand, the classical stability requires $v^2/m^2 \gg 1/\lambda$.
Hence, 
since the bounce action can be of order one or larger
depending on the parameters,  
the 
decay rate can be either large or smaller, respectively.

\section{Quantum nucleation and decay of solitons in external fields
in the thin-defect approximation
}\label{sec:nucleation_wall}

In this section, we give the models 
(sine-Gordon model and complex $\phi^4$ model in the external field) in Subsec.~\ref{sec:model}
and estimate 
nucleation probability of solitons 
in any dimensions $d$  
in terms of 
tensions of solitons and strings (vortices) 
in the thin-defect approximation in Subsec.~\ref{sec:thin-defect-approx}.
We also calculate decay probability of solitons in the external field 
in Subsec.~\ref{sec:decay}.

\subsection{
The models
with external fields}\label{sec:model}
We consider the sine-Gordon model under a constant background field ${\bm B}$ in $3+1$ dimensions, given by
\be
{\cal L}_{\rm IR} = v^2 \left[(\p_\mu\theta)^2 + 2 m^2 (\cos \theta -1) + c {\bm B} \cdot \nabla \theta\right],
\label{eq:Lag_sG}
\ee
where the overall constant $v$ is dimensionful, $m$ is a mass parameter, and the mass dimension of $c$ is $-1$. 
The last term is a total derivative and is a topological term.
This Lagrangian is a simplification of low-energy effective Lagrangians for various interesting systems: chiral magnets with 
easy-axis anisotropy in which the topological term is the DM term, 
or chiral Lagrangian for pions under 
a strong magnetic field (rapid rotation) 
in which the last term originates from 
the CME (CVE). 
Here, we use a notation of ${\bm B}$ for either a magnetic field, a rotation or DM term.
The Hamiltonian reads
\be
{\cal H}_{\rm IR} = v^2 \left[\dot\theta^2 + (\nabla \theta)^2 - 2 m^2 (\cos \theta - 1) - c {\bm B} \cdot \nabla \theta\right].
\ee

Since the last term in Eq.~(\ref{eq:Lag_sG}) is of the first order in the derivative, it does not affect the equation of motion (EOM).
Indeed, the homogeneous configuration $\theta = 2\pi n$ remains the solution of EOM. The energy density is also unchanged from zero.
The soliton solution also remains  to be the same
\be
\theta = 4 \tan^{-1}e^{m \hat {\bm n}\cdot \bm{x}},
\ee
where we have introduced an arbitrary unit vector $\hat{\bm n}$ 
perpendicular to the soliton. 
The solution itself is unchanged, however,
the soliton tension receives a correction from the background field.
Let $\alpha$ be relative angle between $\hat{\bm n}$ and the constant background field ${\bm B}$. Then, the additional energy 
per unit area of the single soliton reads
\be
\delta \sigma
= - 2\pi v^2 c B \cos \alpha,
\ee
where $B = |{\bm B}|$, and we have used the fact that $\theta$ increases by $2\pi$ when one traverses the soliton
along $\hat{\bm n}$.
The net tension of the single soliton reads
\be
\sigma = 16 mv^2 - 2\pi v^2 c B \cos \alpha.
\label{eq:sigma}
\ee
This is minimized when $\alpha = 0$ ($\pi$) for $c > 0$ ($<0$). Namely, the most stable soliton is perpendicular to the
external field ${\bm B}$. 
This implies that the soliton is tensionless at the critical value
\be
B = \frac{8m}{\pi |c|} \equiv B_{\rm c}.
\ee
Moreover, it is negative for $B > B_{\rm c}$.
Therefore, the homogeneous configuration $\theta = 2\pi n$ is no longer ground state, but 
the soliton is the true ground state when $B > B_{\rm c}$.
The multiple solitons are created by increasing $B$, and in general the ground state is a periodic lattice of soliton which
is called the CSL.

The CSLs have been recently studied in various fields. However, most of the previous arguments are static and they do not address 
how the homogeneous ground state is replaced by the soliton when $B$ increases from the value below $B_{\rm c}$ to the one above $B_{\rm c}$. Is the infinitely large soliton suddenly created at the moment of $B=B_{\rm c}$?
This sounds quite unphysical. In order to answer to this elementary question, we study quantum nucleation of the soliton
in this paper.

However, the sine-Gordon model in Eq.~(\ref{eq:Lag_sG}) is not the most suitable for that purpose.
This is because the soliton is topologically stable within the framework of the sine-Gordon model,
and so we cannot discuss neither decay nor nucleation.
Thus, we are naturally guided to a linear sigma model as a UV completion\footnote{From the viewpoint of the microscopic fundamental
theory like QCD, the Lagrangian (\ref{eq:Lag_UV}) should be regarded as a low energy effective theory.} 
by including a massive degree
of freedom (the Higgs mode).
We consider a complex $\phi^4$ model (the axion model 
with the domain wall number one) 
with a constant background field ${\bm B}$:
\be
{\cal L}_{\rm UV} = \left|\p_\mu \phi\right|^2 - \frac{\lambda}{4} \left(|\phi|^2 - v^2\right)^2 + v m^2(\phi + \phi^*) + c {\bm j}\cdot {\bm B}
\label{eq:Lag_UV}
\ee
with
\be
j^\mu = -\frac{i}{2}(\phi^*\p^\mu\phi - \phi\p^\mu\phi^*)  = |\phi|^2 \p^\mu\theta\,,\qquad
\phi = |\phi|e^{i\theta}.
\ee
This reduces back to the sine-Gordon model in Eq.~(\ref{eq:Lag_sG}) in the limit $\lambda \to \infty$. 
Since the last term of Eq.~(\ref{eq:Lag_UV}) does not contribute to EOM, all the solutions for $c=0$ remain to be solutions for $c\neq0$.
Therefore, meta-stable solitons attached by strings should exist in the model (\ref{eq:Lag_UV}) with $c\neq 0$ if
the approximate $U(1)$ condition in Eq.~(\ref{eq:approxU(1)_cond}) is satisfied.

\subsection{Nucleation probability of a soliton 
in the thin-defect approximation}\label{sec:thin-defect-approx}
\begin{figure}[th]
\begin{center}
\includegraphics[width=10cm]{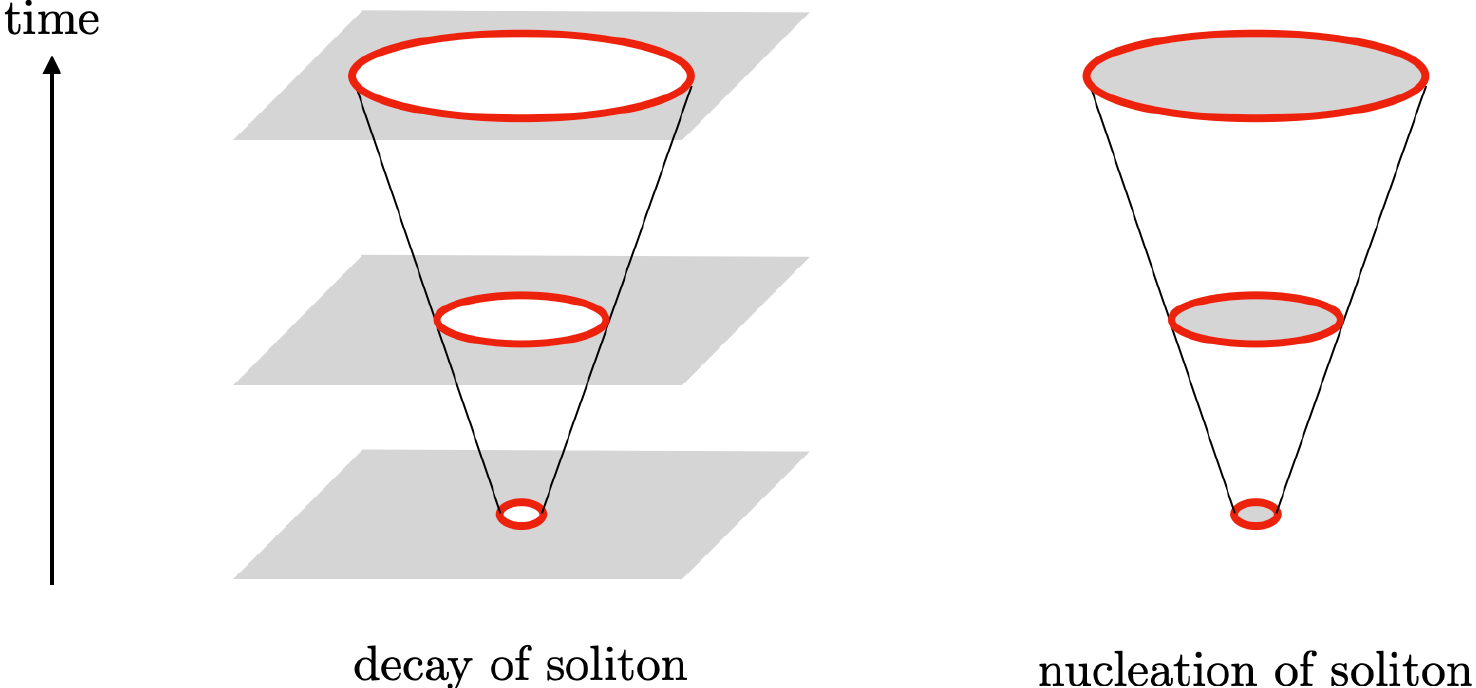}
\caption{The schematic picture of the decay of soliton by nucleating a hole (left) and
the creation of soliton by nucleating a disk soliton (right). 
Grey objects stands for the solitons and
the red objects corresponds to the strings.}
\label{fig:decay_vs_nucleation}
\end{center}
\end{figure}

Now let us take $B$ which is ssmaller than $B_c$. The ground state should be the homogeneous configuration.
Then, we increase $B$ to any value above $B_c$ instantaneously. 
The ground state should be solitonic in this case.
To estimate the probability of nucleation of a soliton, we reverse the arguments about the soliton decay in Sec.~\ref{sec:decay_wall}
in which the nucleation probability of {\it a hole on the domain wall} 
was calculated, see Fig.~\ref{fig:decay_vs_nucleation}.

Here, we consider {\it a disk of a soliton bounded by a string loop 
in the homogeneous vacuum}.
Let us consider the spatial dimension $d$, 
and later we will set $d=2,3$. 
In the thin-defect approximation, 
the bounce action  reads
\be
	S(R) = {\rm vol}(S^{d-1}) R^{d-1} \mu + {\rm vol}(B^{d}) R^d \sigma,
\label{eq:bounce_action}
\ee
where $R$ is the radius of the soliton,  
and 
${\rm vol}(S^{d-1}) R^{d-1}$ and  ${\rm vol}(B^{d})$
are volumes of the unit hypersphere and hyperball, given by
\be
{\rm vol}(S^{d-1}) = \frac{2 \pi ^{\frac{d}{2}}}{\Gamma \left(\frac{d}{2}\right)},\qquad
{\rm vol}(B^{d}) = \frac{\pi ^{d/2}}{\Gamma \left(\frac{d+2}{2}\right)},
\ee
respectively. 
Note that the string tension $\mu$ is not logarithmically divergent but a finite constant  
in the presence of the $U(1)$ breaking term (the third term in Eq.~(\ref{eq:Lag_UV})).
See Appendix \ref{sec:app2} for some details. 

Clearly, $\mu$ is always positive.
In contrast, the soliton tension $\sigma$ given in Eq.~(\ref{eq:sigma}) can be either positive or negative.
In the absence of the topological term ($c=0$),
$\sigma$ is positive, 
and $S$ has no stationary points except for $R=0$. 
Namely, quantum nucleation of the disk is prohibited.

However, the situation drastically changes in the presence of the topological term, $c \neq 0$, because the soliton tension 
can be negative $\sigma < 0$ for $B > B_{\rm c}$. Then, a nontrivial stationary point exists at
\be
R = \frac{d-1}{d}\frac{{\rm vol}(S^{d-1})}{{\rm vol}(B^{d})}\frac{\mu}{(-\sigma)}.
\ee
and the nucleation probability can be calculated as
\be
P_{\text{nucleation}} = A e^{-S},\quad 
S = \frac{(d-1)^{d-1}}{d^d}\frac{\left[{\rm vol}(S^{d-1})\right]^d}{\left[{\rm vol}(B^{d})\right]^{d-1}}\frac{\mu^d}{(-\sigma)^{d-1}}.
\label{eq:S0}
\ee
Since
$-\sigma$
given in Eq.~(\ref{eq:sigma})
is maximized at $\alpha = 0$ ($\pi$) for $c > 0$ ($<0$), the bounce action is minimized
there with the negative soliton tension
\be
\sigma_\perp 
= 16 mv^2 \left(1 -\frac{B}{B_{\rm c}}\right) < 0.
\ee 
Therefore, the nucleation probability is maximized for the soliton
perpendicular to ${\bm B}$. Once the disk perpendicular to ${\bm B}$ is nucleated, it rapidly expands.
The thin-defect approximation is justified for $R \gg \delta_{\rm dw}$.
This can be rewritten as $\mu/(-\sigma_\perp) \gg 1/m$.

\subsection{Decay probability of a soliton in external fields 
in the thin-defect approximation} 
\label{sec:decay}

Here we consider quantum decay of a soliton in the external field.
Consider an infinitely large flat soliton perpendicular to the external field
under the external field larger than $B_{\rm c}$.
The bounce action of a hole on a soliton 
can be written
in the thin-defect limit as
\be
S(R) = {\rm vol}(S^{d-1}) R^{d-1} \mu - {\rm vol}(B^{d}) R^d \sigma_0\left(1 - \frac{B}{B_{\rm c}}\right),
\ee
with $\sigma_0 = 16mv^2$. Since the second term is positive for $B> B_{\rm c}$, 
the bounce action does not have stationary point, and therefore the decay 
is {\it prohibited}.
We should emphasized that 
the soliton which is metastable for $B=0$ is 
completely stable for $B> B_{\rm c}$.

On the other hand, when we instantaneously decrease $B$ below $B_{\rm c}$, the stationary point of the bounce action
appears at
\be
R = \frac{B_{\rm c}}{B_{\rm c}-B} \frac{d-1}{d}\frac{{\rm vol}(S^{d-1})}{{\rm vol}(B^{d})}\frac{\mu}{\sigma_0},
\ee
and the value of the action reads
\be
S = 
\left(\frac{B_{\rm c}}{B_{\rm c}-B}\right)^{d-1}
\frac{(d-1)^{d-1}}{d^d}\frac{\left[{\rm vol}(S^{d-1})\right]^d}{\left[{\rm vol}(B^{d})\right]^{d-1}}\frac{\mu^d}{\sigma_0^{d-1}}.
\ee
Comparing this with the bounce action without external field, we find
\be
S\big|_{B=0} < S\big|_{0<B<B_{\rm c}}.
\ee 
This implies that the decay rate of the soliton is suppressed by the external field.
As the external field $B$ increases toward $B_{\rm c}$ from below, the action diverges and the quantum decay is strongly suppressed.
 
\section{Numerical simulations for quantum nucleation of solitons} \label{sec:numerical}
In this section, we numerically calculate nucleation probability 
of solitons.
In Subsec.~\ref{eq:preliminary} we rewrite the Lagrangian 
in terms of dimensionless variables.
In Subsecs.~\ref{sec:nucleation2+1} and \ref{sec:nucleation3+1}, 
we calculate nucleation probabilities of solitons 
by numerically constructing bounce solutions  
in 2+1 and 3+1 dimensions, respectively.

\subsection{Preliminary}\label{eq:preliminary}

A great benefit of considering the linear sigma model in Eq.~(\ref{eq:Lag_UV}) is that we can treat the soliton and strings
as  regular objects of finite sizes. With them at hand we can go beyond the thin-defect limit.

We will numerically solve EOM of Lagrangian in Eq.~(\ref{eq:Lag_UV}).
To this end, it is useful to rewrite Eq.~(\ref{eq:Lag_UV}) in terms of the dimensionless variables
\be
\tilde x^\mu = m x, \quad \tilde \phi = v^{-1} \phi,\quad \tilde \lambda = \frac{m_h^2}{m^2},\quad
\tilde {\bm B} = m^{-1} c{\bm B}.
\ee
Then, we have the Lagrangian
\be
{\cal L}_{\rm UV} 
= m^2 v^2 \left[
|\tilde\p_\mu \tilde\phi|^2 - \frac{\tilde\lambda}{4} \left(|\tilde\phi|^2 - 1\right)^2 + \tilde\phi + \tilde\phi^* + \tilde {\bm j}\cdot \tilde{\bm B}
\right]
\ee
where 
$\tilde \lambda$ is the unique parameter characterizing solutions. For the (meta-)stable solitons and strings to exist, we need
to assume $\tilde\lambda \gg 1$ corresponding to the condition in Eq.~(\ref{eq:approxU(1)_cond}).
For concreteness, we will assume that the soliton is perpendicular to the $z$-axis. Therefore, the last term
in the bracket can be written as
\be
\tilde {\bm j}\cdot \tilde{\bm B} = \tilde B \tilde j_z \cos \alpha.
\ee
The Hamiltonian reads
\be
{\cal H}_{\rm UV} = m^2 v^2 \tilde {\cal H}_{\rm UV},\quad
\tilde {\cal H}_{\rm UV} = 
\left|\tilde\nabla \tilde \phi_z\right|^2 
+ \frac{\tilde\lambda}{4} \left(|\tilde\phi|^2 - 1\right)^2
- (\tilde\phi + \tilde\phi^*) -  \tilde j_z \tilde B \cos \alpha.
\label{eq:dimless_H}
\ee

\subsection{Quantum nucleation of a soliton in $2+1$ dimensions}
\label{sec:nucleation2+1}

Here, we investigate nucleation of solitons in $2+1$ dimensions, 
in which the soliton is a linear object and the vortex is a particle object.
The $2+1$ dimensional version of Eq.~(\ref{eq:bounce_action}) with $d=2$ is
\be
S = 2\pi R \mu + \pi R^2 \sigma.
\label{eq:S_2+1}
\ee
Its extremum is given by Eq.~(\ref{eq:S0}) for $d=2$,
\be
R_0 = \frac{\mu}{-\sigma},\qquad S_0 = \frac{\pi \mu^2}{-\sigma}.
\label{eq:R0S0_2d}
\ee
Note that we have $[v]=\frac{1}{2}$, $[\lambda] = 1$, and $[m]=1$ in $2+1$ dimensions.
We will compare this analytic formula in the thin-defect limit with numerical simulations for the soliton with finite thickness.

Once we obtain a numerical solution for a soliton attached by a vortex and an anti-vortex at its both ends,
we can measure the dimensionless radius $\tilde R$ (a half length) of the soliton and evaluate 
the dimensionless total energy $\tilde{\cal E}$ by
\be
\tilde {\cal E}(\tilde R) = \int d^2\tilde x\  \tilde {\cal H}_{\rm UV}(\tilde R).
\ee
From this, we can evaluate the bounce action through the following formula
\be
S 
= \int d^3x\, {\cal H}_{\rm UV} 
= \frac{v^2}{m} \int d^3\tilde x\, \tilde{\cal H}_{\rm UV} 
= 
\alpha_1 \frac{v^2}{m} \int_0^{\tilde R} d\tilde r~\tilde{\cal E}(\tilde r),
\label{eq:EtoSd=2}
\ee
with a constant $\alpha_1 = \pi$ (see Appendix \ref{sec:app1} for a derivation).
To understand the formula quickly, let us substitute the energy formula $\tilde{\cal E}(\tilde R) = 2\tilde\mu + 2\tilde R \tilde \sigma$ 
with $\tilde \mu = \mu/v^2$ and $\tilde \sigma = \sigma/(mv^2)$
in the thin-defect limit
(the soliton of the length $2\tilde R$ with two vortices). We easily find that 
the bounce action in Eq.~(\ref{eq:S_2+1}) is correctly reproduced.

Note that the formula in Eq.~(\ref{eq:EtoSd=2}) is valid only for constant $\mu$. If $\mu$ depended on $R$ logarithmically as
the usual global vortex, we cannot use Eq.~(\ref{eq:EtoSd=2}).
In Appendix \ref{sec:app2}, we show our numerical solution 
in which the profile of the scalar field 
exponentially converges to the VEV in the asymptotic region,
in contrast to the usual global vortex without any domain walls 
for which the profiles polynomially approaches to the VEV.

By differentiating $S$ by $R$, we have
\be
\frac{dS}{dR} 
= \alpha_1 \frac{v^2}{m}\frac{d\tilde R}{dR} \tilde{\cal E}(\tilde R)
= \alpha_1 v^2 \tilde{\cal E}(\tilde R).
\ee
The extremum of $S$ is then identified with the zero of ${\cal E}$:
\be
\frac{dS}{dR} = 0
\quad\Leftrightarrow\quad
\tilde {\cal E}(\tilde R) = 0.
\ee

The remaining task is constructing suitable numerical configurations with a soliton bounded by a vortex and an anti-vortex.
To this end, we use the standard relaxation scheme. Our method consists of two steps. Firstly, we take a product 
ansatz of a pair of a vortex and an anti-vortex separated at distance $2\tilde R_{\rm ini}$ as an initial configuration of the relaxation. 
At this stage,
we fix the positions of the vortices. Then, the straight soliton of the length $2\tilde R_{\rm ini}$ is generated and the configuration converges quite soon.
We use this convergent configuration as the initial configuration for the second relaxation, 
in which 
we do not fix the vortex positions. 
During the second relaxation process, the vortices 
approach to each other due to
the soliton tension, and eventually annihilate each other.
We repeatedly measure the distance $2\tilde R$ of the vortices and compute $\tilde{\cal E}(\tilde R)$.

\begin{figure}[t]
\begin{center}
\includegraphics[width=15cm]{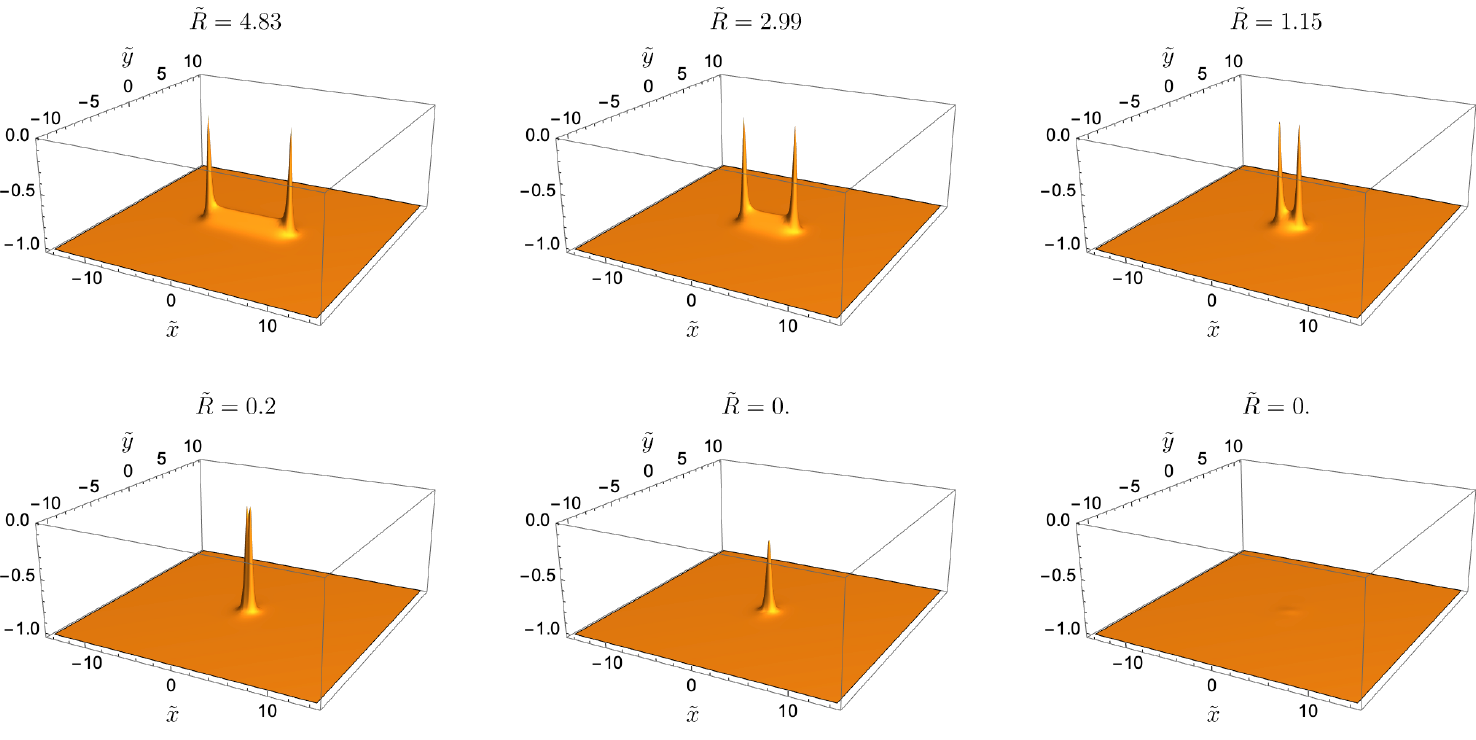}
\caption{The soliton attached by a vortex and anti-vortex in $2+1$ dimensions. The negative amplitude $(-|\tilde\phi|)$ is plotted.}
\label{fig:amplitude_2dim}
\end{center}
\end{figure}
To be concrete, 
we take $\tilde \lambda = 100$ which is large enough for the soliton and vortices to be classically metastable. 
The amplitudes $(-|\tilde\phi|)$ for several different separations are shown in Fig.~\ref{fig:amplitude_2dim}. 
The two peaks correspond to the vortex and anti-vortex while the linear object stretching between them is the soliton
which is visible only for a large $\tilde R$.
We evaluate $\tilde{\cal E}$ with three different values of $\tilde B \cos\alpha = \{1,1/2,1/4\}$ in Eq.~(\ref{eq:dimless_H}).

\begin{figure}[th]
\begin{center}
\includegraphics[width=15cm]{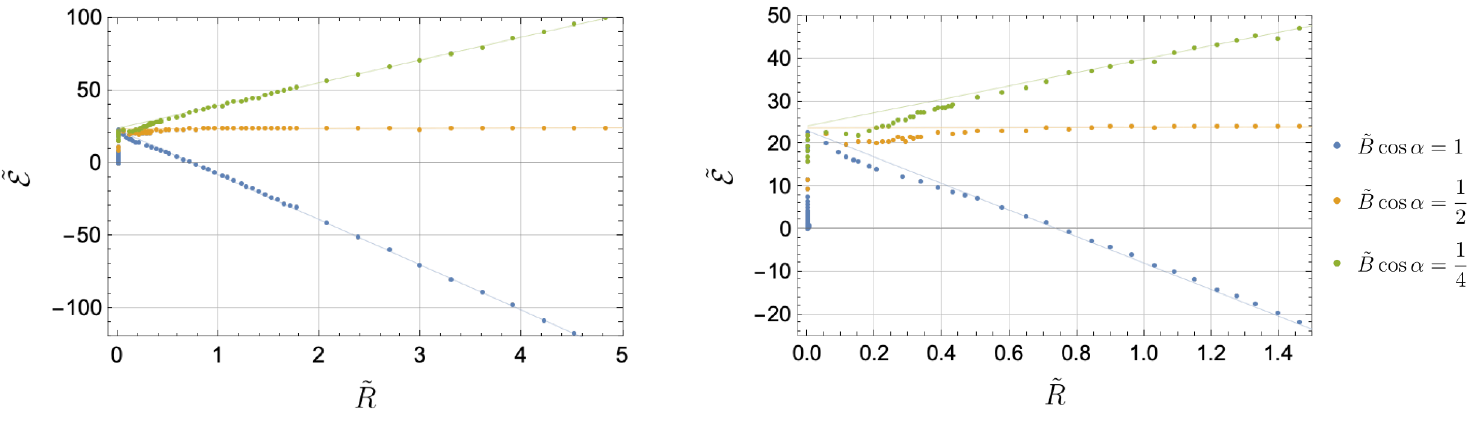}
\caption{The dimensionless total energy $\tilde{\cal E}$ is plotted as the function of the dimensionless radius $\tilde R$.
Three different choices of $\tilde B \cos\alpha = \{1,1/2,1/4\}$ 
corresponding to $B \geq B_c$, $B = B_c$ and $B \leq B_c$, respectively, are shown. The right panel is the zoom-up around $\tilde R=0$ of the left one.}
\label{fig:EvsR_2dim}
\end{center}
\end{figure}
The results are shown in Fig.~\ref{fig:EvsR_2dim}. The datas are well fitted by a linear function in the large $\tilde R$ region
\be
\tilde{\cal E} = 2\tilde R a  + 2 b,
\label{eq:fit_2dim}
\ee
which should be compared with the thin-defect limit $\tilde{\cal E} = 2\tilde R \tilde\sigma  + 2 \tilde\mu$.
The coefficient $a$ can be either positive or negative because it is related to the soliton tension
which depends on $\tilde B \cos\alpha$.
On the other hand, the constant $b$ should be insensitive on $\tilde B \cos \alpha$ because it should be identified with the vortex tension
which is independent on the background field.
Indeed, our numerical solutions show that the three lines almost meet at $\tilde R = 0$ in Fig.~\ref{fig:EvsR_2dim}.
The coefficients read numerically are shown in Table.\ref{tab:ab_2dim}.
Thus we numerically determine the tensions $\tilde\mu$ and $\tilde\sigma$. Importantly, $\tilde\mu$ is a constant as we mentioned above,
see also Appendix \ref{sec:app2}.
\begin{table}
\begin{center}
\begin{tabular}{c|cc}
$\tilde B \cos\alpha~(d=2)$ & $a(\sim \tilde \sigma)$ & $b(\sim \tilde \mu)$ \\
\hline
$1$ & $- 15.6$ & $11.6$\\
$1/2$ & $0.07$ & $12.0$\\
$1/4$ & $7.85$ & $12.2$
\end{tabular}
\qquad
\begin{tabular}{c|ccc}
$\tilde B \cos\alpha~(d=3)$ & $a(\sim \tilde \sigma)$ & $b(\sim \tilde \mu)$ & $c$ \\
\hline
$1$   & $-15.3$ & $11.8$ & $45.6$\\
$1/2$ & $-0.15$ & $12.7$ & $28.4$\\
$1/4$ & $7.78$ & $12.3$ & $24.3$
\end{tabular}
\caption{The numerically obtained coefficients in Eqs.~(\ref{eq:fit_2dim}) and (\ref{eq:fit_3d}).}
\label{tab:ab_2dim}
\end{center}
\end{table}

Among the three different choices, only $\tilde B \cos \alpha = 1$ leads to the negative soliton tension, 
corresponding to the case of $B \geq B_c$.
The stationary point is $\tilde R_0 \simeq 0.75$, and the value of the bounce action is about $\tilde S \simeq b\tilde R \simeq 9.0$.
Hence, the nucleation probability 
can be estimated as
\be
P_{\rm nucleation} = A \exp\left(-\alpha_1\frac{v^2}{m} \times 9.0\right).
\ee
Note that the numerically determined values $(\tilde R_0,\tilde \sigma, \tilde\mu) = (0.75, - 15.6, 11.6)$ is 
consistent with the analytic formula $\tilde R_0 = \tilde \mu/(-\tilde\sigma)$ for the thin-defect limit.

\subsection{Quantum nucleation of a soliton in $3+1$ dimensions}
\label{sec:nucleation3+1}

Next, we numerically investigate quantum nucleation of a disk soliton bounded by a string loop in $3+1$ dimensions.
By putting $d=3$ in Eq.~(\ref{eq:bounce_action}), the bounce action of the thin-defect limit is given by
\be
S = \pi R^2 \mu + \frac{4\pi}{3} R^3 \sigma,
\label{eq:S_3+1}
\ee
and its extremum is given by Eq.~(\ref{eq:S0}) for $d=3$,
\be
 R_0 = \frac{2\mu}{-\sigma},\qquad 
 S_0 = \frac{16\pi \mu^3}{3\sigma^2}.
\label{eq:R0S0_d3}
\ee

The numerical procedures we adopt in this subsection are 
the same as those in the previous subsection
except for differences due to the spacial dimensions.
We numerically evaluate 
the dimensionless mass $\tilde{\cal E}$ for a disk of a soliton of the radius $R$
\be
\tilde{\cal E}(\tilde R) = \int d^3\tilde x~\tilde{\cal H}_{\rm UV},
\ee
where the dimensionless Hamiltonian $\tilde{\cal H}_{\rm UV}$ is given in Eq.~(\ref{eq:dimless_H}).
Then, we evaluate the bounce action $S$ by the following formula
\be
S(R) 
= \int d^{4}x\, {\cal H}_{\rm UV} 
= \frac{v^2}{m^{2}} \int d^{4}\tilde x\, \tilde{\cal H}_{\rm UV} 
=
\alpha_2 \frac{v^2}{m^2}  \int_0^{\tilde R} d\tilde r~\tilde{\cal E}(\tilde r).
\label{eq:S_d=3}
\ee
The constant factor is $\alpha_2 = 4$ whose derivation is given in Appendix \ref{sec:app1}.
Again, we should note that this formula is valid for the constant $\mu$.
As a quick check of the formula, one can reproduce Eq.~(\ref{eq:S_3+1}) by substituting
$\tilde{\cal E} = 2\pi\tilde R\tilde\mu + \pi \tilde R^2 \tilde \sigma$.

\begin{figure}[ht]
\begin{center}
\includegraphics[width=15cm]{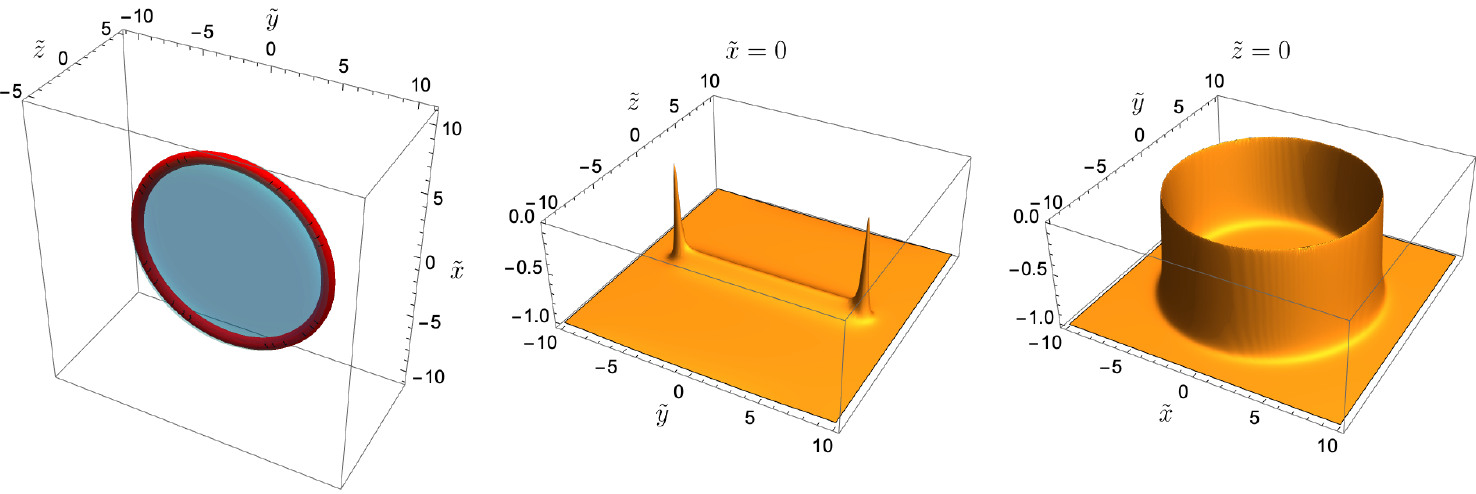}
\caption{The three-dimensional disk soliton perpendicular to the $z$-axis generated from the two-dimensional
linear soliton. In the left panel the red part shows the region for $|\tilde\phi|<0.9$, and 
the blurred-blue part corresponds to the region for ${\rm Re}[\tilde\phi] < 0.4$.
The middle and right panels show $-|\tilde\phi|$ on the $\tilde x = 0$ and $\tilde z = 0$ cross sections, respectively.}
\label{fig:disk_initial}
\end{center}
\end{figure}
As before we take relatively large value $\tilde \lambda = 100$ to assure the classical stability of the solitons
and strings. In order to prepare a disk shape soliton, we recycle the numerical configuration of 
the linear soliton attached by two vortices in $2+1$ dimensions.
Set the two-dimensional linear soliton along the $x$-axis on the $xz$-plane, and let $\phi_{\rm 2d}(x,z)$ be the 
corresponding field configuration. Then, a three-dimensional disk soliton perpendicular to the $z$-axis can be obtained by
rotating it around the $z$-axis, namely
$\phi_{\rm 3d}(x,y,z) = \phi_{\rm 2d}(x\cos\theta+y\sin\theta,z)$ with $\tan \theta =y/x$, see Fig.~\ref{fig:disk_initial}.

Having this as an initial configuration, we evolve it by a standard relaxation method.
The disk soliton shrinks as the relaxation process proceeds. We measure the radius and the mass,
so that we determine the function $\tilde{\cal E}(\tilde R)$.
Finally, we calculate the bounce action $S(R)$ by plugging it to the formula in Eq.~(\ref{eq:S_d=3}).
\begin{figure}[ht]
\begin{center}
\includegraphics[width=15cm]{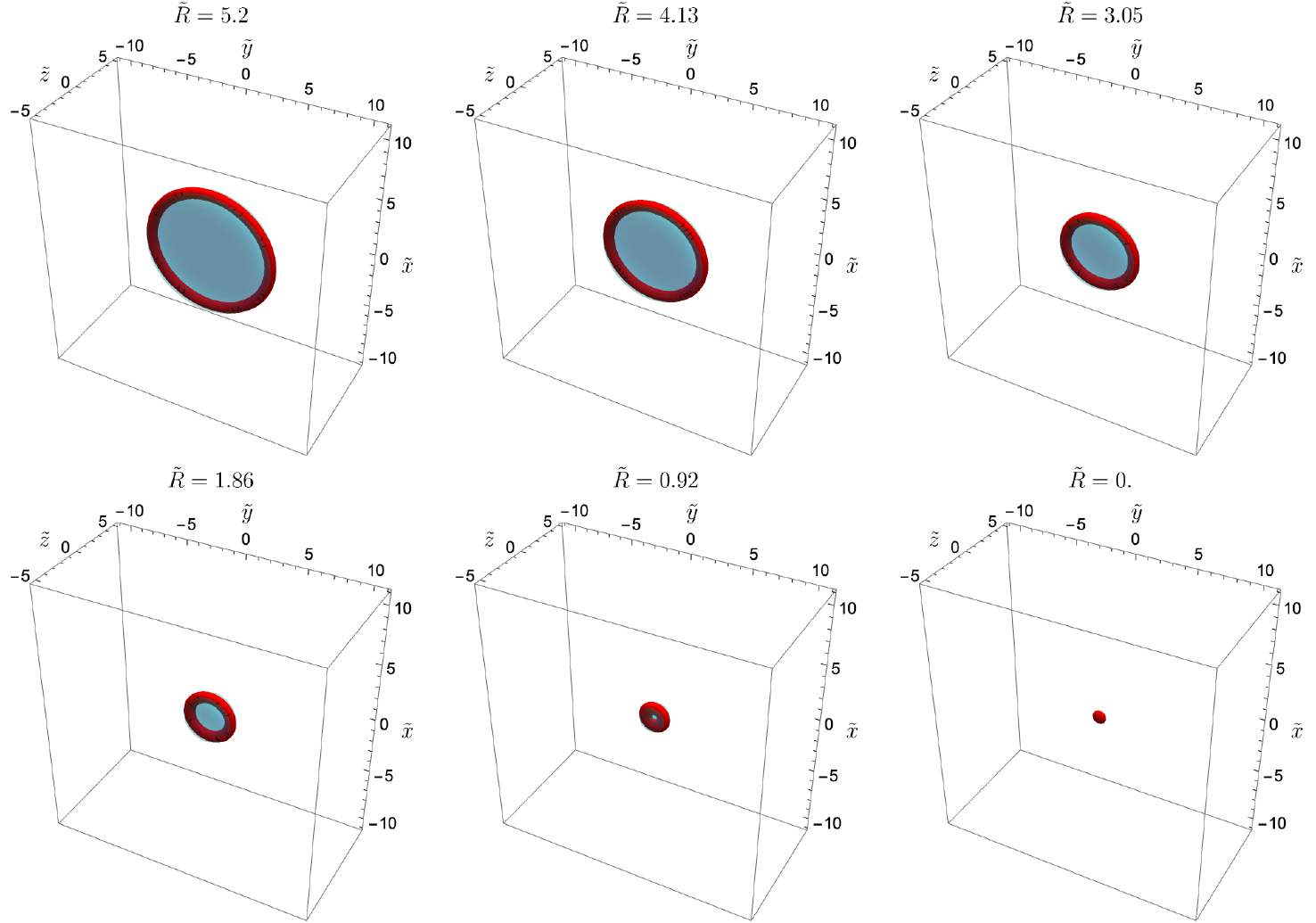}
\caption{The disk soliton bounded by a ring string.
The red part shows the region for $|\tilde\phi|<0.9$, and 
the blurred-blue part corresponds to the region for ${\rm Re}[\tilde\phi] < 0.4$.}
\label{fig:rings_3d}
\end{center}
\end{figure}
The time evolution of the disk soliton under the relaxation process is shown in Fig.~\ref{fig:rings_3d}.
The disk is initially large, and we can clearly observe a circular closed string  (red part: $|\tilde\phi|<0.9$) 
and a disk soliton (blurred-blue part: ${\rm Re}[\tilde\phi] < 0.4$).
We determine the radius $\tilde R$ of the ring by seeking the point where $|\tilde\phi| = 0$.
Note that the energy does not immediately vanishes when $\tilde R$ reaches zero. This is a finite width effect which
is missed in the thin-defect limit. Since the soliton and string are regular objects with finite sizes, 
a remnant of energy still exists and it gradually decays and finally disappears.

\begin{figure}[t]
\begin{center}
\includegraphics[width=15cm]{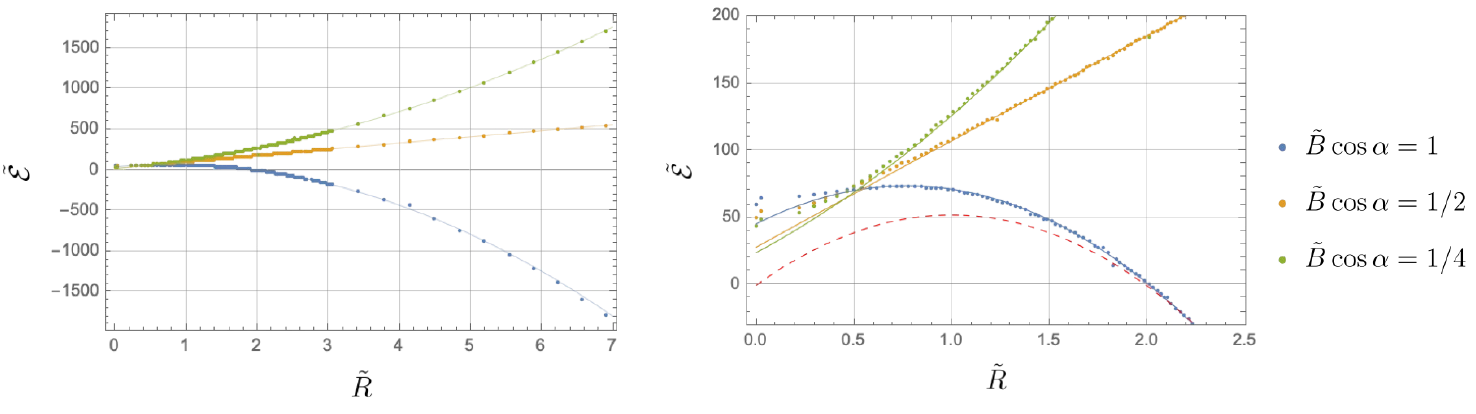}
\caption{The dimensionless total energy $\tilde{\cal E}$ is plotted as the function of the dimensionless radius $\tilde R$ for $d=3$.
We set $\tilde B \cos\alpha = \{1,1/2,1/4\}$. 
The right panel is the zoom-up near $\tilde R=0$ of the left one.}
\label{fig:EvsR_3dim}
\end{center}
\end{figure}
The numerical results for $\tilde B \cos\alpha = \{1,1/2,1/4\}$ are shown in Fig.~\ref{fig:EvsR_3dim}.
The fact that the energy of $\tilde B \cos\alpha = 1$ is negative for large $\tilde R$ indicates that the soliton tension is negative.
The zero of $\tilde{\cal E}$ is found as $\tilde R \simeq 2$ which should correspond to the extremum point of the bounce action.
The energy of $\tilde B \cos\alpha = 1/2$ is approximately a linear function of $\tilde R$ implying the soliton tension vanishes.
That of $\tilde B \cos\alpha = 1/4$ grows faster than linear functions, implying that the soliton tension is positive.
These are well fitted by
\be
\tilde{\cal E} = \pi \tilde R^2 a + 2\pi \tilde R b + c.
\label{eq:fit_3d}
\ee
The values of these coefficients are shown in Table \ref{tab:ab_2dim}.
The values of $a$ ($\sim \tilde \sigma$) and $b$ ($\sim\tilde \mu$)
are consistent between the left ($d=2$) and right ($d=3$) tables in Table~\ref{tab:ab_2dim}.
The last term $c$ corresponds to the remnant energy at $\tilde R=0$. This is absent in the thin-defect limit. Hence,
the bounce action for the finite size soliton is slightly larger than the one 
at the thin-defect limit. 

By using the fit in Eq.~(\ref{eq:fit_3d}), we can evaluate the action for $\tilde B \cos\alpha = 1$ as
\be
S = \alpha_2 \frac{v^2}{m^2} \int_0^{\tilde R_0=2} d\tilde r~\tilde{\cal E}(\tilde r) \simeq  111 \alpha_2 \frac{v^2}{m^2}.
\ee
Thus, the nucleation probability can be obtained as 
\be
P_{\rm nucleation} = A \exp\left(- 111 \alpha_2 \frac{v^2}{m^2}\right).
\label{eq:nucleation-rate}
\ee

Before closing this section, let us examine the relevance of 
the remnant energy found above.
If we fit the numerical data by ignoring the remnant energy with forcing $c=0$, then we find $a=-16.7$ and $b=16.7$,
see the red-dashed curve in Fig.~\ref{fig:EvsR_3dim}.
The value of integration for the constrained fit is $\int_0^{\tilde R_0=2} d\tilde r~\tilde{\cal E} \simeq 69$,
so that the nucleation probability is slightly increased. Here, 
$a=-16.7$ is still consistent with the one obtained in the $d=2$ case, whereas
$b=16.7$ shows a relatively large discrepancy from $b$ in the $d=2$ case.
Hence, the string tension $b$ is not correctly captured by the constrained fit. Thus, we conclude that 
the remnant energy $c$ 
included in the unconstrained fit is not a sort of artifacts of the numerical simulation but it should be
considered as a real finite width effect.

In conclusion, we have succeeded in numerically evaluating the bounce action for the soliton bounded by the string
with finite thickness. The finite width effect has been found and it slightly reduces the nucleation probability 
compared with the thin-defect limit.

\section{Summary and discussion}\label{sec:summary}

We have proposed quantum nucleation of topological solitons 
through quantum tunneling, 
as a novel mechanism for formation of topological solitons. 
We have discussed chiral solitons in a complex $\phi^4$ model (an axion model) with a topological term, 
which is a low-energy theory of  
chiral magnets with an easy-plane anisotropy  
and QCD at finite density under strong magnetic field or rapid rotation. 
First, we have estimated the creation probability analytically 
in terms of tensions of string (vortex) and soliton 
in the thin-defect approximation in any dimension. 
Second,  
we  have performed numerical simulations 
in $2+1$ and $3+1$ dimensions by
the relaxation (gradient flow) method,
and have obtained creation probabilities. 
We have found a good agreement between the thin-defect approximation 
and direct numerical simulation in $2+1$ dimensions  
and have found, in $3+1$ dimensions, 
a difference between them at short distances at the subleading order, which can be interpreted as the remnant energy.

We have considered the complex $\phi^4$ model as an UV theory 
for the sine-Gordon model at IR limit appearing in various context. 
Sine-Gordon solitons are almost insensitive to UV, 
but different UV theories give different structures of strings 
(vortices). 
However, creation probabilities will be insensitive to such details.

\begin{figure}[ht]
\begin{center}
\includegraphics[width=15cm]{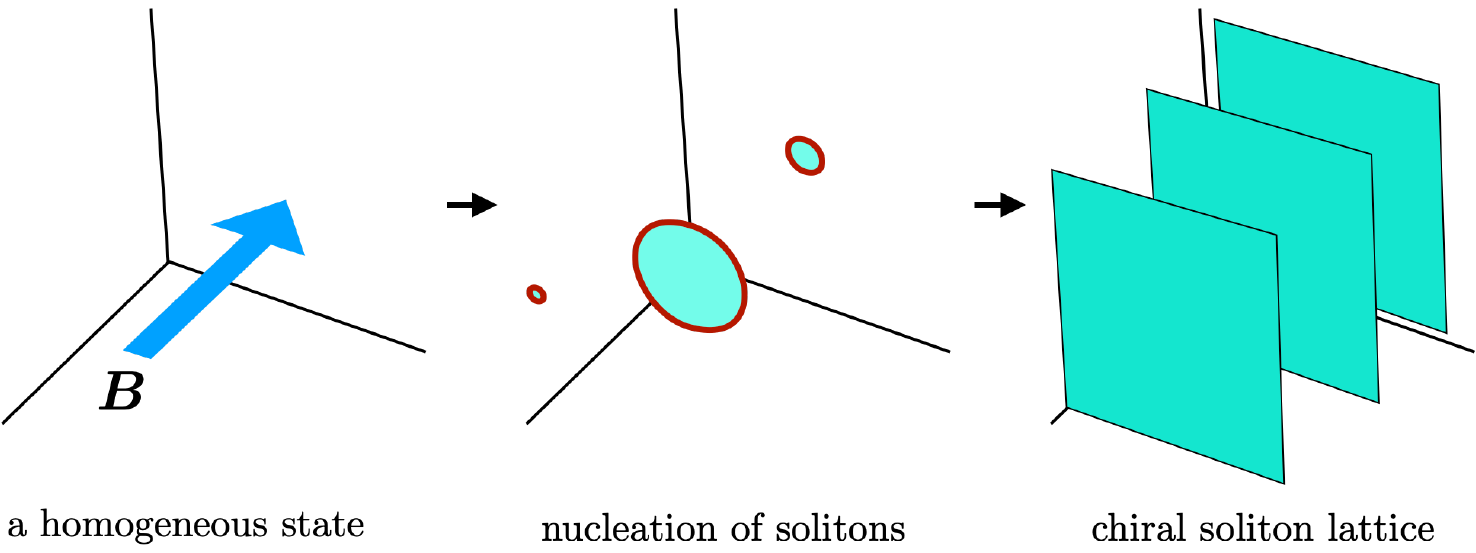}
\caption{Nucleation of solitons and formation of a CSL.}
\label{fig:CSL}
\end{center}
\end{figure}

In this study, we have estimated 
the nucleation probability in the vacuum 
where there are no solitons.
In the case with 
the external field 
above the critical value, 
the true ground state is a CSL.
Formation of the CSL in the homogeneous vacuum should occur in the following process.
Let us turn on the external field $B > B_{\rm c}$ in 
the homogeneous vacuum. 
Initially, disk solitons of the critical radius $R_0$ 
in Eq.~(\ref{eq:R0S0_d3})
are nucleated all over with the creation rate 
in Eq.~(\ref{eq:nucleation-rate}).
They rapidly expand 
as in the right panel of Fig.~\ref{fig:decay_vs_nucleation}, 
growing up to infinitely large solitons. 
These solitons repel each other and thus 
adjust intersoliton distances to minimize the total energy,
thereby eventually forming into a CSL 
as schematically shown in Fig.~\ref{fig:CSL}.
Of course, this is a rough sketch and we need more 
detailed analysis of 
dynamical process. 
We also need to calculate 
nucleation probabilities of solitons not in
the homogeneous vacuum but also in inhomogeneous soliton backgrounds. 
For instance, once the CSL ground state is formed, 
nucleation probabilities of solitons should be zero 
in such a background. 
If we instantaneously increase (decrease) the external field 
in the CSL ground state, 
the number density of solitons 
 should be decreased (increased).
We thus need nucleation (decay) rates of 
solitons in the CSL background, 
which remain as a future work.

In this paper, we have considered the Abelian sine-Gordon model 
for simplicity. 
On the other hand,
it was found in Ref.~\cite{Eto:2021gyy} 
that 
in the case of two-flavor baryonic matter under rotation, 
non-Abelian solitons with non-Abelian ${\mathbb C}P^{N-1}$ moduli 
\cite{Nitta:2014rxa,Eto:2015uqa}
are also present 
in the ground state of QCD in a certain parameter region.
In this case, a non-Abelian soliton is bounded by 
a non-Abelian global string \cite{Nitta:2014rxa, Eto:2021nle}.
In such a case, the creation probability may depend on 
the dimension $N$ of the moduli as numerical factor.

In $2+1$ dimensions, 
the pseudo-NG mode is mapped to an electromagnetic field 
 under a duality,
while vortices are mapped to charged particles. 
With nonzero mass $m$, particles are confined 
by electric fluxes.
In this duality,  the topological term will be mapped to 
a constant electric field. 
It remains as a future problem to study nucleation probabilities 
in terms of the duality.
In $2+1$ dimensions, there is a BKT transition
at finite temperature. 
It is interesting to discuss 
 whether there is any conflict between 
 quantum nucleation and the BKT transition.

Note added:
While this paper is being completed, we were informed that 
the authors of Ref.~\cite{Nishimura} was preparing 
a draft which may have some overlap with our work. 


\section*{Acknowledgements}
This work is supported in part by JSPS Grant-in-Aid for Scientific Research (KAKENHI Grant No. JP22H01221). 
The work of M. E. is supported in part by the JSPS Grant-in-Aid for Scientific Research KAKENHI Grant No. JP19K03839 and
the MEXT KAKENHI Grant-in-Aid for Scientific Research on Innovative Areas ``Discrete Geometric Analysis for Materials Design'' 
No. JP17H06462 from the MEXT of Japan.


\begin{appendix}

\section{Asymptotic behavior of the scalar field 
of a vortex attached by a soliton}
\label{sec:app2}

Here, we show 
a numerical solution with some more details to demonstrate that 
the amplitude of $\phi$ for a vortex attached by 
a soliton decays exponentially fast, 
in contrast to the slow-tail with a negative power of 
isolated global vortices without solitons. 
For this purpose, we use a configuration of 
a vortex and an anti-vortex connected by a soliton 
discussed in Sec.~\ref{sec:nucleation2+1}, see Fig.~\ref{fig:amplitude_2dim}. 
When the vortex and anti-vortex are well separated compared 
with the width of the soliton, 
the configuration around the vortex 
is close to that of a single vortex attached by a soliton.
\begin{figure}[ht]
\begin{center}
\includegraphics[width=15cm]{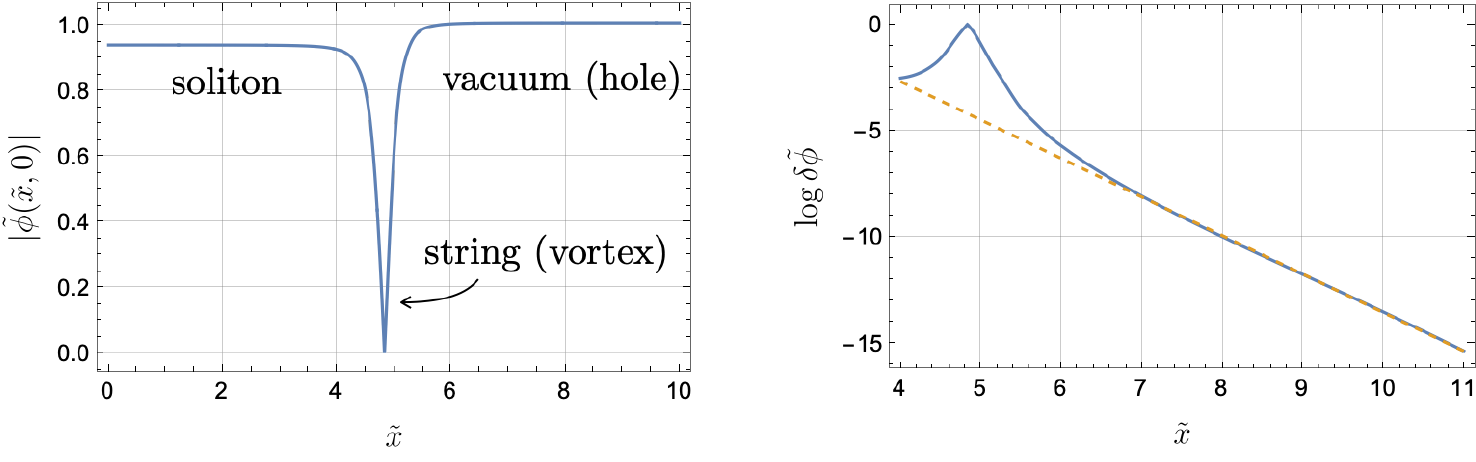}
\caption{The left panel shows $\tilde\phi(\tilde x,0)$ for the configuration shown in the top-left
panel of Fig.~\ref{fig:amplitude_2dim}, and the right panel shows asymptotic behavior of the same solution.}
\label{fig:asym}
\end{center}
\end{figure} 
The left panel of Fig.~\ref{fig:asym} shows 
a scalar field profile 
$\tilde\phi(\tilde x,0)$ at a cross section $y=0$ of
the configuration shown in the top-left
panel of Fig.~\ref{fig:amplitude_2dim} with $\tilde R=4.83$. 
One can see that apart from the vortex core,
the scalar field $\tilde \phi$ quickly converges 
to a constant in the vacuum. 
To confirm that the asymptotic behavior is an exponential tail, we show 
$\log\delta\tilde\phi = \log (\tilde v - |\tilde\phi|)$
in the right panel of Fig.~\ref{fig:asym}. We numerically fit the asymptotic tail and the result is
 $\delta\tilde\phi(\tilde x,0) \propto e^{-1.8\tilde x}$. 
 Thus, the amplitude converges exponentially fast to the VEV.

We can confirm the same behaviors 
in any directions from the vortex center except for 
the directions of the soliton within its width. 
This fact implies 
 that the vortex tension is finite as in Eq.~(\ref{eq:mu}), 
 in contrast to an isolated global vortex 
 whose tension is logarithmically divergent 
 as in Eq.~(\ref{eq:global_string}). 
 Physically, 
the logarithmically divergent behavior (of an isolated global vortex) is replaced by 
 the soliton tension.
 This point was missed in the literature 
 \cite{Preskill:1992ck,Vilenkin:2000jqa} 
 in which the vortex tension was assumed to be 
 logarithmically divergent even when the vortex is attached by a soliton.
 This is crucial to evaluate  
 bounce actions in the thin-defect limit
 and likewise the decay rates 
 and nucleation probabilities  
 of topological solitons.


\section{A derivation of Eqs.~(\ref{eq:EtoSd=2}) and (\ref{eq:S_d=3})}
\label{sec:app1}

In this appendix, we give a derivation of Eqs.~(\ref{eq:EtoSd=2}) and (\ref{eq:S_d=3}) for the spatial dimension $d=2,3$.
However, we will consider generic $d$ below.
Since the number of codimensions of strings(vortices) is 2 and that of  solitons is 1, a soliton is a $(d-1)$-dimensional ball 
$B^{d-1}$ while the string wrapping the soliton is a $(d-2)$-dimensional sphere $S^{d-2}$.
The volumes of unit $d$-sphere and $d$-ball are given by
\be
{\rm vol}(S^{d-1}) = \frac{2 \pi ^{\frac{d}{2}}}{\Gamma \left(\frac{d}{2}\right)},\qquad
{\rm vol}(B^{d}) = \frac{\pi ^{d/2}}{\Gamma \left(\frac{d+2}{2}\right)}.
\ee
Therefore, the mass in the thin-defect limit reads
\be
\tilde {\cal E}(\tilde R) 
= \int d^{d}\tilde x\, \tilde{\cal H}_{\rm UV} 
= {\rm vol}(S^{d-2}) \tilde R^{d-2} \tilde \mu + {\rm vol}(B^{d-1})  \tilde R^{d-1} \tilde \sigma,
\ee
with the dimensionless tensions of string and soliton
\be
\tilde \mu = \mu/v^2,\quad
\tilde \sigma = \sigma/(mv^2),
\ee
respectively. 
On the other hand, the string world-volume is $S^{d-1}$ and the soliton world-volume is $B^{d-1}$
for the bounce action, and we have
\be
\tilde S(\tilde R)  \equiv \int d^{d+1}\tilde x\, \tilde{\cal H}_{\rm UV} 
={\rm vol}(S^{d-1}) \tilde R^{d-1}\tilde \mu + {\rm vol}(B^{d}) \tilde R^{d} \tilde \sigma.
\ee
Integrating $\tilde {\cal E}(\tilde R)$, we find
\be
\int^{\tilde R}_0 d\tilde r~\tilde {\cal E}(\tilde r) 
&=& \frac{1}{d-1}{\rm vol}(S^{d-2}) \tilde R^{d-1} \mu + \frac{1}{d}{\rm vol}(B^{d-1})  \tilde R^{d} \sigma \nonumber\\
&=& \left(\frac{1}{d-1}\frac{{\rm vol}(S^{d-2})}{{\rm vol}(S^{d-1})}\right) {\rm vol}(S^{d-1})\tilde R^{d-1} \mu 
+ \left(\frac{1}{d}\frac{{\rm vol}(B^{d-1})}{{\rm vol}(B^{d})}\right) {\rm vol}(B^{d}) \tilde R^{d} \sigma. \;\;\;
\ee
Note that the two ratios are identical as
\be
\alpha_{d-1} \equiv (d-1) \frac{{\rm vol}(S^{d-1})}{{\rm vol}(S^{d-2})}
= d \frac{{\rm vol}(B^{d})}{{\rm vol}(B^{d-1})},
\ee
and therefore we have
\be
\tilde S(\tilde R) = \alpha_{d-1} \int^{\tilde R}_0 d\tilde r~\tilde {\cal E}(\tilde r).
\ee
We assume that this formula is valid not only for the thin-defect limit but also the case that the defects have regular sizes.
We have used $\alpha_1 = \pi$ and $\alpha_2 = 4$ in the text.

\end{appendix}

\bibliographystyle{jhep}
\bibliography{references}

\end{document}